\documentclass[useAMS]{mn2e}
\usepackage{psfig}
\newif\ifAMStwofonts
\def\la{\;
\raise0.3ex\hbox{$<$\kern-0.75em\raise-1.1ex\hbox{$\sim$}}\; }
\def\ga{\;
\raise0.3ex\hbox{$>$\kern-0.75em\raise-1.1ex\hbox{$\sim$}}\; }
\newcommand{\zabs}{$z_{\rm abs}\,$}
\newcommand{\hh}{H$_2\,\,$}
\begin{document}
\title[$m_{\rm p}/m_{\rm e}$ non-variability]
{A new constraint on cosmological variability of 
the proton-to-electron mass ratio\thanks{
Based on public data released from UVES Commissioning 
at the VLT Kueyen telescope, ESO, Paranal, Chile}
}
\author[S.~A.~Levshakov et al.]
{S.~A.~Levshakov,$^1$\thanks{On leave from the
Ioffe Physico-Technical Institute, Russian Academy of Sciences, St.~Petersburg}
M.~Dessauges-Zavadsky,$^2$\thanks{Affiliated to
Observatoire de Gen\`eve, CH--1290 Sauverny, Switzerland}
S.~D'Odorico,$^2$ and P.~Molaro$^3$\\
$^1$
Division of Theoretical Astrophysics, National Astronomical Observatory, 
Tokyo 181-8588, Japan\\
$^2$
European Southern Observatory, Karl-Schwarzschild-Str. 2,
D-85748 Garching bei M\"unchen, Germany\\
$^3$
Osservatorio Astronomico di Trieste,
Via G.B. Tiepolo 11, I-34131 Trieste, Italy
}
\date{Accepted .
      Received ;
      in original form }
\pagerange{\pageref{firstpage}--\pageref{lastpage}}
\pubyear{2001}
\maketitle
\label{firstpage}
\begin{abstract}
Exotic cosmologies predict variability of the fundamental physical constants over the
cosmic time. Using the VLT/UVES high resolution spectra of
the quasar Q0347--3819 and unblended electronic-vibrational-rotational
lines of the \hh molecule identified at \zabs = 3.025 we test possible changes in the
proton-to-electron mass ratio $\mu_0 = m_{\rm p}/m_{\rm e}$
over the period of $\sim 11$ Gyr.
We obtained a new constraint on the
time-averaged variation rate of $\mu_0$ of
$|\dot{\mu}/\mu_0| < 5\times10^{-15}$ yr$^{-1}$ ($1\sigma$ c.l.). 
The estimated $1\sigma$
uncertainty interval of the $| \Delta \mu/\mu_0 |$ ratio of about
0.004\% implies that since the time when the \hh spectrum was formed at
$z_{\rm abs} = 3.025$,
$\mu_0$ has not changed by more than a few thousands of a percent.
\end{abstract}

\section{Introduction}

\addtocounter{footnote}{3}

Kaluza-Klein (KK) type models 
(Super-symmetric Grand Unification Theory, Super string models, etc.)
unify gravity with other fundamental forces.
These models predict
variations of the fundamental physical constants
over the cosmological evolution (for a review, see, e.g.,
Okun' 1991).
Variations of the coupling constants
of strong and electroweak interactions 
would affect the masses of the
elementary particles in a way which 
depends on the adopted scenario for the expanding universe. 
The model parameters of these theories can be constrained by observations. 
One possibility to test values of the physical constants 
at different cosmological epochs is to
study high resolution spectra of extragalactic objects
(Savedoff 1956). 

Such analysis has been recently carried
out by a number of authors (see, e.g., Varshalovich, Potekhin 
\& Ivanchik 2000, and references cited therein), 
and a possible variation of the
fine-structure constant, 
$\alpha_0 = e^2/\hbar\,c$, at a level of
$\Delta \alpha/\alpha_0 = (-0.72\pm0.18)\times10^{-5}$
was announced by Webb et al. (2001) who
analyzed fine-splitting lines in quasar spectra.

The first restriction on the variability of 
the proton-to-electron mass ratio\footnote{The present value of
the proton-to-electron mass ratio is 
$\mu_0 = 1836.1526670$ (Mohr \& Taylor 2000).},
$\mu_0 = m_{\rm p}/m_{\rm e}$, stemming from
quasar spectra $| \Delta \mu/\mu_0| < 0.13$ ($1\sigma$ c.l.)
was obtained by Pagel (1977) who compared the observational
wavelengths of H\,{\sc i} and metals as previously proposed 
by Thompson (1975).
However, the derived upper limit on $\mu_0$ 
depends on the assumption that all elements have the same
fractional ionization ratios and trace the same volume elements
along the respective line of sight. This assumption
may not be true in general
especially for QSO absorbers where
complex absorption-line profiles are observed at high spectral resolution.

The  proton-to-electron mass ratio can be estimated
more accurately from high redshift
molecular hydrogen systems. With some modifications, such measurements
were performed for the $z_{\rm abs} = 2.811$ H$_2$ system from the spectrum
of PKS 0528--250 by Foltz, Chaffe \& Black (1988), Varshalovich \& Levshakov (1993),
Cowie \& Songaila (1995), and by Potekhin et al. (1998) who set the most
stringent limit of $|\Delta \mu/\mu_0|< 1.8\times10^{-4}$ ($1\sigma$ c.l.). 

In this paper, we present a new upper limit on the variation rate
of the proton-to-electron mass ratio obtained from the analysis of
a new H$_2$ system found at $z_{\rm abs} = 3.025$ toward the quasar
Q0347--3819. 

\begin{figure*}
\vspace{0.0cm}
\hspace{0.0cm}\psfig{figure=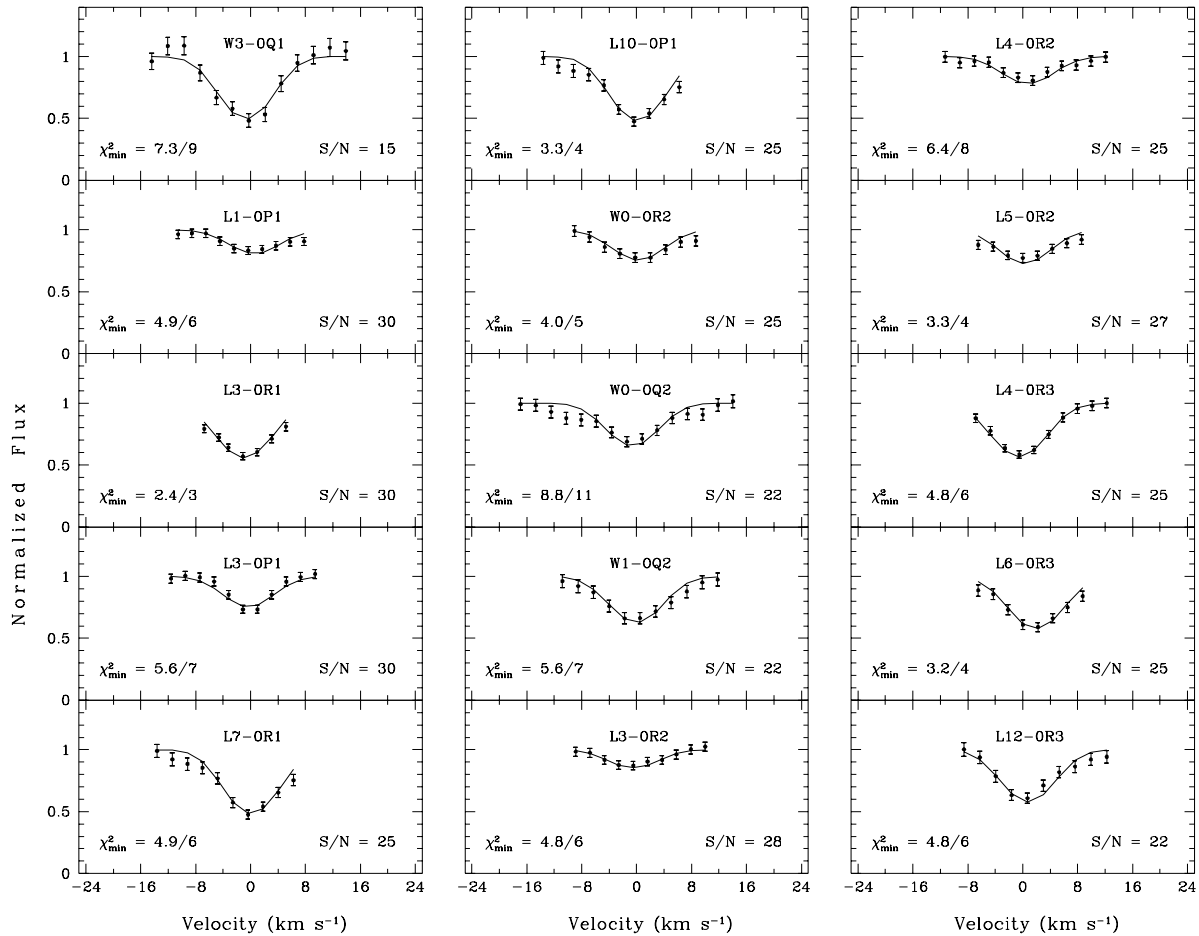,height=18.0cm,width=18.7cm}
\vspace{-4.5cm}
\caption{H$_2$ absorption features associated with the $z_{\rm abs} = 3.025$ damped 
Ly$\alpha$ system toward Q0347--3819 (renormalized intensities shown by dots with $1\sigma$
error bars were calculated for the mean values of the local continuum deviations
$\Delta C/C$ listed in Table ~1). The zero radial velocity is fixed at
$z_{{\rm H}_2} = 3.024895$. Smooth curves are the synthetic H$_2$ profiles found by the
least-squares procedure as described in the text. The best ($\chi^2$/degree of freedom)
values and the mean $S/N$ per resolution element
are shown in the corresponding panels.
}
\label{fig1}
\end{figure*}

\section{Observations}

High resolution spectra of the quasar
Q0347--3819 were obtained during UVES commissioning at
the VLT 8.2m ESO telescope and are described in detail by
D'Odorico, Dessauges-Zavadsky \& Molaro (2001)  
and Levshakov et al. (2002, LDDM hereinafter).
The spectrum resolution 
is FWHM $\simeq 7.0$ km~s$^{-1}$ and 5.7 km~s$^{-1}$
in the UV and near-IR ranges, respectively. The
signal-to-noise ratio per resolution element in the UV range is
$S/N \sim 10-40$. 

We identified more than 80 \hh molecular lines 
in a damped Ly$\alpha$ system (DLA hereinafter)\footnote{
The damped Ly$\alpha$ systems are
believed to originate in the intervening galaxies or proto-galaxies
located at cosmological distances (e.g. Wolfe et al. 1995).}
at \zabs = 3.025 toward Q0347--3819 (LDDM).
Some of them are not suitable for further analysis due to
H\,{\sc i} Ly$\alpha$ forest contamination. However, 
we selected 15 unblended \hh lines (shown in Fig.~1) which
provide the most accurate line center measurements to set an upper limit on
possible changes of $\mu_0$.

We would like to emphasize that our analysis of the \zabs = 3.025 DLA
has shown that
all the \hh line profiles can be adequately described with
a unique value of $z_{{\rm H}_2} = 3.024855\pm0.000005$\footnote{
Based on the H$_2$ laboratory wavelengths given in Abgrall \& Roueff (1989).} 
which implies that no assumption
on the variability of $\mu_0$ is needed or {\it statistically justified}. 
Observations show that the ratio
$\Delta \mu/\mu_0 \equiv (\mu_z - \mu_0)/\mu_0$
is zero. The analysis presented below has the objective to estimate at what accuracy
$\Delta \mu/\mu_0 = 0$.

\begin{table*}
\centering
\begin{minipage}{140mm}
\caption{H$_2$ lines at \zabs = 3.025 toward Q0347--3819 and sensitivity
coefficients ${\cal K}$.} 
\begin{tabular}{c l r @{$\pm$} l cc r @{.} l r @{.} l} \hline
$J$ & Line & \multicolumn{2}{c}{$\lambda^a_0$, \AA} 
& $\Delta C/C^b$,\,\% &
$\lambda_{\rm obs}$, \AA & \multicolumn{2}{c}{$z$} & 
\multicolumn{2}{c}{${\cal K}$} \\ 
(1) & (2) & \multicolumn{2}{c}{(3)} & (4) & (5) & \multicolumn{2}{c}{(6)} & 
\multicolumn{2}{c}{(7)} \\  \hline 
\noalign{\smallskip}
1 & W3-0Q & 947.4218&0.0005 & $0.9\pm0.2$ & $3813.2653\pm0.0041$ & 3&024887(5)$^d$ 
& 0&0217427(8)\\
 & L1-0P & 1094.0522&$0.0051^c$ & $0.8\pm0.1$ & $4403.4575\pm0.0077$ & 3&02491(2)
& $-0$&0023282(1)\\
 & L3-0R & 1063.4603&$0.0001^c$ & $0.8\pm0.1$ & $4280.3051\pm0.0030$ & 3&024885(3)
& 0&0112526(4)\\
 & L3-0P & 1064.6056&0.0005 & $0.9\pm0.2$ & $4284.9249\pm0.0033$ & 3&024894(4) 
& 0&0102682(4)\\
 & L7-0R & 1013.4412&0.0020 & $0.8\pm0.1$ & $4078.9785\pm0.0025$ & 3&024898(9) 
& 0&03050(1)\\
 & L10-0P&  982.8340&0.0006 & $0.8\pm0.1$ & $3955.8049\pm0.0051$ & 3&024896(6) 
& 0&04054(6)\\
\noalign{\smallskip}
2 & W0-0R & 1009.0233&0.0007 & $0.8\pm0.1$ & $4061.2194\pm0.0061$ & 3&024901(7) 
& $-0$&0050567(7)\\
 & W0-0Q & 1010.9380&0.0001 & $0.8\pm0.1$ & $4068.9088\pm0.0053$ & 3&024885(6) 
& $-0$&0068461(2)\\
 & W1-0Q & 987.9744&0.0020  & $0.9\pm0.1$ & $3976.4943\pm0.0049$ & 3&02490(1) 
& 0&0039207(2)\\
 & L3-0R & 1064.9935&0.0009 & $1.0\pm0.2$ & $4286.4755\pm0.0098$ & 3&02488(1) 
& 0&0097740(5)\\
 & L4-0R & 1051.4981&$0.0004^c$ & $0.8\pm0.1$ & $4232.1793\pm0.0085$ & 3&024904(8)
& 0&015220(1)\\
 & L5-0R & 1038.6855&0.0032 & $0.8\pm0.1$ & $4180.6048\pm0.0070$ & 3&02490(1) 
& 0&020209(3)\\
\noalign{\smallskip}
3 & L4-0R & 1053.9770&0.0011 & $0.8\pm0.1$ & $4242.1419\pm0.0028$ & 3&024890(5) 
& 0&012837(2)\\
  & L6-0R & 1028.9832&0.0016 & $0.8\pm0.1$ & $4141.5758\pm0.0031$ & 3&024921(7)  
& 0&022332(7)\\
  & L12-0R& 967.6752&0.0021 & $0.8\pm0.1$  & $3894.7974\pm0.0039$ & 3&02490(1) 
& 0&0440(2)\\ \hline
\noalign{\smallskip}
\multicolumn{10}{l}{Notes: $^a$\,listed values are from Abgrall et al. (1993a, 1993b);
$^b$\,the local continuum deviation with $1\sigma$ }\\
\multicolumn{10}{l}{error; $^c$ the error is estimated from the comparison between the data in
Abgrall \& Roueff (1989) and}\\
\multicolumn{10}{l}{Abgrall et al. (1993a, 1993b); 
$^d$\,$1\sigma$ standard deviations are shown in parenthesis (last digits after a} \\
\multicolumn{10}{l}{decimal point), 
e.g. 3.024887(5) means $3.024887\pm0.000005$. }\\
\end{tabular}
\end{minipage}
\end{table*}

\section{Data analysis and results}

For measurements of the absorption line centers in QSO spectra, 
there are three principal sources of statistical
errors  caused both by the data reduction procedure and by 
statistical fluctuations in the recorded counts. First, all echelle data must
be rebinned to constant wavelength bins in order to combine different spectra 
and hence to increase the
signal-to-noise ratio. Such resampling changes statistical properties of the noise
and introduces correlations between the data point values. The correlation coefficient
of about +0.8 was estimated by LDDM for the UVES data in question. 
This means that the classical
$\chi^2$ test is {\it no longer applicable} 
to the rebinned spectra, because the measurements are not
independent. The second error, 
usually important for lines observed in the Ly$\alpha$ forest,
arises from the uncertainty of the overall continuum level and its shape
in the vicinities of individual spectral features. 
The third error is connected with the
finite number of photons counted. 
The individual intensity points within the interval of integration
can fluctuate producing distortions in the line shape.

There are, of course, other sources of errors in the measurements of the
precise wavelengths $\lambda_{\rm obs}$ of the observed H$_2$ lines.
The values of $\lambda_{\rm obs}$ may be affected by errors in the 
wavelength calibration and  by the thermal shifts.  
The observed wavelengths are not significantly affected by
errors in the correction of the wavelength scale to the heliocentric and vacuum values.
Typically, these errors do not exceed a few m~s$^{-1}$ (e.g.
Edl\'en 1953).
The quasar spectrum is, however, constantly shifted with temperature by 0.37 pixel/K.
Given the approximate pixel size (0.02 \AA) at 4000 \AA\,, we can expect 
the shift of at most
0.3 km~s$^{-1}$ in case the Th calibration spectrum was taken a few hours after the
science exposures and the difference in temperature was 0.5~K. The thermal
shift should be more or less constant for all lines, so it affects only the
assumed redshift. The values of $\lambda_{\rm obs}$ are mainly affected by errors
in the wavelength calibration. We found that the rms of this 
systematic error over the 
total range  covered by the blue spectrum is 
$\sigma_{\rm syst} = \pm0.0016$ \AA. 
This corresponds at 4000 \AA\, to 0.12 km~s$^{-1}$. 

The continuum placement in the UV portion of the Q0347--3819 spectrum 
was determined by using `continuum windows' 
in the Ly$\alpha$ forest which were fitted by a low
order polynomial. 
The accuracy  $\delta_c$ of the local continuum (which may deviate from
the common continuum in the Ly$\alpha$ forest region) was estimated during
the fitting procedure described below.
 
For a given \hh line, the number of pixels involved in the analysis 
corresponds to the number
of points in the line profile shown in Fig.~1. 
To measure the line centers we used a method
that matches the observed profiles with the synthesized ones to estimate 
the set of model parameters.
Our previous analysis of the line profiles of different elements 
identified in the \zabs = 3.025
H$_2$-bearing cloud has shown that both 
low ions and \hh have the same simple velocity structure
of the main component -- a narrow symmetrical core with  a common broadening parameter
$b_{{\rm H}_2} = 2.80\pm0.45$ km~s$^{-1}$ (LDDM). 
Thus in our present study we also used a simple 
one-component model with four free parameters: the center, the width, the intensity
of the absorption line  and the local continuum displacement 
$\delta_c = \Delta C/C$ (this technique
was successfully used in the study 
of metal lines in the Ly$\alpha$ forest by Molaro et al. 2001).
The set of initial parameters then was adjusted until a satisfactory fit
could be achieved. 
The objective function was augmented with the penalty function in the
form [cf. Eqs. (13) and (14) in LDDM]:
\begin{equation}
\psi = \left( \frac{b - b_{{\rm H}_2}}{\sigma_{b_{{\rm H}_2}}} \right)^2\; ,
\label{eq:E3}
\end{equation}
where $b_{{\rm H}_2} = 2.80$ km~s$^{-1}$ and $\sigma_{b_{{\rm H}_2}} = 0.45$ 
km~s$^{-1}$.

\begin{figure}
\vspace{0.0cm}
\hspace{0.0cm}\psfig{figure=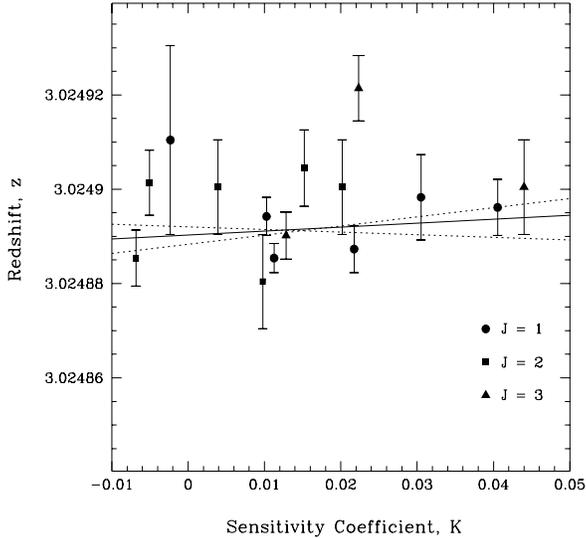,height=8.0cm,width=8.0cm}
\vspace{-0.5cm}
\caption{Relation between the redshift 
values $z_i$ calculated for individual H$_2$ features shown
in Fig.~1 and their sensitivity coefficients ${\cal K}_i$. 
The solid line is the linear regression:
$z_i = \bar{z} + \kappa ({\cal K}_i - \bar{\cal K})$ 
with $\kappa = (1 + \bar{z})\Delta \mu/\mu_0$.
The dotted lines representing the $1\sigma$ deviations 
from the slope $\kappa$ of the best linear regression
demonstrate that $\Delta \mu/\mu_0 \equiv 0$ at the level $\sim 3\times10^{-5}$.
}
\label{fig2}
\end{figure}

To evaluate statistical errors, 
Monte Carlo analysis was performed in the same way as described in
Appendix A in LDDM. The results of these calculations are presented in columns (4) and (5) of
Table~1. It is seen that for the selected H$_2$ lines
$\delta_c \simeq 1$\%.
This means that the H$_2$ lines involved in the analysis are not significantly
disturbed by the Ly$\alpha$ forest absorption.

In this approach
the central wavelengths $\lambda_{\rm obs}$
can be measured with an accuracy exceeded the average pixel size. 
As expected, the weaker absorption features
have lower accuracy compared with the stronger lines. 
The corresponding redshifts for individual \hh lines are shown in column (6) with
the $1\sigma$ uncertainties including $\sigma_{\rm syst}$.
In these calculations we used the \hh laboratory wavelengths 
$\lambda_0$ from 
Abgrall et al. (1993a, 1993b) which show slightly different $\lambda_0$ values as
compared with their previous publication (Abgrall \& Roueff 1989). 
Therefore, as a consequence the new
values of $z_{{\rm H}_2}$ discussed below are shifted with respect to 
those presented in LDDM.

\begin{figure}
\vspace{0.0cm}
\hspace{0.0cm}\psfig{figure=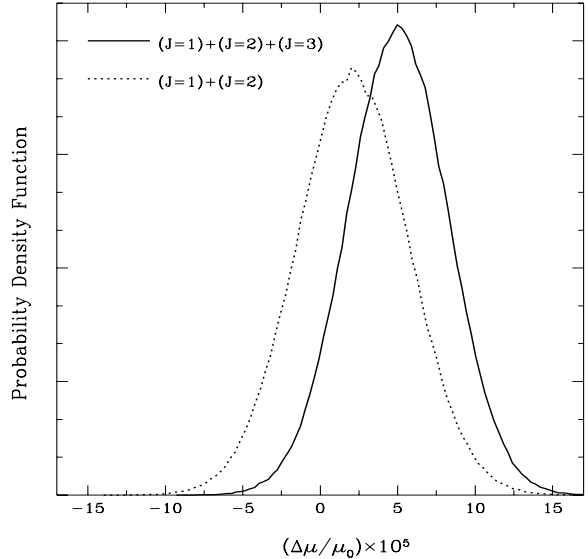,height=8.0cm,width=8.0cm}
\vspace{-0.5cm}
\caption{
Monte Carlo simulations of the probability density functions (p.d.f.)
of
$\Delta \mu/\mu_0$ for two samples: $(i)$ the H$_2$ lines 
arising from the low rotational levels with
$J=1,2,3$ (the solid curve), and $(ii)$ the H$_2$ lines 
arising from the low rotational levels with
$J=1,2$ (the dashed curve). The difference between p.d.f. is caused
by gradual shift in the radial velocity as $J$ increases.
}
\label{fig3}
\end{figure}

With the obtained $z$ values and their standard deviations we are able to constrain possible
changes of $\mu_0$. Indeed,
if the ratio $\Delta \mu/\mu_0 \equiv (\mu_z - \mu_0)/\mu_0$
is not zero,
then for any two \hh lines with rest frame wavelengths
$\lambda_{i,0}$ and $\lambda_{j,0}$ the ratio
\begin{equation}
\frac{\lambda_{i,z}/\lambda_{j,z}}{\lambda_{i,0}/\lambda_{j,0}}
\simeq 1 + ({\cal K}_i - {\cal K}_j)\,\Delta\mu/\mu_0
\label{eq:E4}
\end{equation}
would deviate from unity, where $\lambda_{i,z}$ and $\lambda_{j,z}$ are
the corresponding line centers measured in a quasar spectrum.

In equation (\ref{eq:E4}), the so-called sensitivity coefficients ${\cal K}$ determine
the sensitivity of \hh wavelengths to the variation of the
proton-to-electron mass ratio. These coefficients have been calculated
by Varshalovich \& Levshakov (1993) and in a different manner by
Varshalovich \& Potekhin (1995). Both methods give
results in good agreement.
The most recent version of the procedure to calculate ${\cal K}$
and $\Delta \mu/\mu_0$ from QSO spectra is described in detail
by Potekhin et al. (1998). Following this procedure we calculated
the sensitivity coefficients which are listed in column (7) of Table~1.
The accuracy of the ${\cal K}$ values was also estimated 
from coefficients $Y_{mn}$ (listed in Table~1 in Potekhin et al.)
using the method of error propagation ($Y_{mn}$ values were considered to be accurate
to $k$ decimal places and their rounding errors were set to $0.5\times10^{-k}$). 
It should be noted 
that although transitions between the excited states of the H$_2$ molecule have
higher wavelength-to-mass sensitivity coefficients, their accuracy is much lower as
compared with the low-$J$ transitions.
In linear approximation 
\begin{equation}
z_i = \bar{z} + \kappa ({\cal K}_i - \bar{\cal K})\; , 
\label{eq:E5}
\end{equation}
where
$\kappa = (1 + \bar{z})\Delta\mu/\mu_0$ with $\bar{z}$ and $\bar{\cal K}$ being
the mean redshift and the mean sensitivity coefficient, respectively.

The linear regression in the form (\ref{eq:E5}) was firstly calculated for the
complete sample of the \hh lines (Table~1) where transitions from the $J = 1, 2$ and 3
levels are combined. The obtained result
$(\Delta \mu/\mu_0)_{J=1+2+3} = (5.0\pm3.2)\times10^{-5}$ (which is consistent with
the value independently found by  Ivanchik et al. 2002)
shows a ``possible variation'' of $\mu_0$ at the $1.6\sigma$ level.
However, if we consider \hh transitions from individual $J$ levels, then
the weighted mean redshifts reveal a {\it gradual shift in the radial velocity} for
features arising from progressively higher rotational levels~:
$z^{J=1}_{{\rm H}_2} = 3.024890(2)$, $z^{J=2}_{{\rm H}_2} = 3.024895(3)$, and
$z^{J=3}_{{\rm H}_2} = 3.024904(4)$\footnote{
The numbers shown in parenthesis are
the $1\sigma$ standard deviations in last digits after a decimal point.}.
The H$_2$ lines with changing profiles and small velocity shifts as $J$ increases
were also observed in our Galaxy in the direction of $\zeta$ Ori~A by Jenkins \&
Peimbert (1997) who suggested that these H$_2$ lines 
are formed in {\it different} zones of a postshock gas. 

Albeit the $z^{J=1}_{{\rm H}_2}$ and $z^{J=2}_{{\rm H}_2}$ values are consistent
within $1\sigma$ intervals, the difference between $z^{J=1}_{{\rm H}_2}$ and
$z^{J=3}_{{\rm H}_2}$ is essential and equals $1.0\pm0.3$ km~s$^{-1}$.
If we now exclude from the regression analysis levels with $J=3$, then
$(\Delta \mu/\mu_0)_{J=1+2} = (2.1\pm3.6)\times10^{-5}$. 
(The use of samples with \hh lines arising from the same rotational levels
would be more reasonable to estimate $\Delta\mu/\mu_0$, 
but in our case the sample size is rather small and we have to combine the
$J=1$ and $J=2$ levels to increase accuracy).
This linear regression is shown by the solid line in Fig.~2 while two dashed lines
correspond to the $1\sigma$ deviations of the slope parameter $\kappa$. 

We also calculated the probability density functions of $\Delta \mu/\mu_0$ 
for two samples of the \hh lines  (with $J = 1, 2, 3$, and $J = 1, 2$)
using statistical Monte Carlo simulations which suggest that the errors
$\sigma_z$ and $\sigma_{\cal K}$  are normally distributed around the mean
values of $z$ and ${\cal K}$ with the dispersions equal to their
probable errors listed in Table ~1.
The result is presented in Fig.~3. It is clearly seen that small changes in the
radial velocity with increasing $J$
cause the difference between these two probability density
functions and thus may mimic a shift in $\Delta \mu/\mu_0$.

From these calculations we find
$|\Delta \mu/\mu_0| < 5.7\times10^{-5}$ ($1\sigma$) which
is about three times stronger
as compared with the value estimated by Potekhin et al. (1998).

Both constraints on the variation of the $m_{\rm p}/m_{\rm e}$ ratio
are in good agreement with the limit on the variability of the
product of the fine-structure constant, nuclear $g$ factor of the
proton and the masses of the electron and proton
$\Delta {\rm ln}\,(\alpha^2\,g_{\rm p}\,m_{\rm e}/m_{\rm p}) =
(1.2 \pm 1.8)\times10^{-4}$ which was set by de Bruyn, O'Dea \&
Baum (1996) from the measurements of the redshifts of the 
H\,{\sc i} 21 cm
line and the optical resonance absorption lines
observed at \zabs = 3.38 in the DLA toward Q0201+113.
The same order of magnitude restrictions to
$\Delta {\rm ln}\,(\alpha^2\,g_{\rm p}\,m_{\rm e}/m_{\rm p})$
were found from other DLAs detected at \zabs = 0.524 (Q0235+164),
0.692 (3C286), 1.944 (Q1157+014), and 2.038
(Q0458-020) (see Table~4 in de Bruyn et al. 1996), 
and even stronger limit of
$(0.7\pm1.1)\times10^{-5}$ at \zabs = 1.776 (Q1331+170) was set
by Cowie \& Songaila (1995).
Comparison of H\,{\sc i} 21 cm and molecular absorption
(CO, $^{13}$CO, C$^{18}$O, CS, HCO$^+$, and HCN) also
yields very tight constrains on the ratio
$|\Delta (\alpha^2\,g_{\rm p})/(\alpha^2\,g_{\rm p})_0| \la 0.7\times10^{-5}$
at \zabs = 0.2467 and 0.6847 toward Q1413+135 and Q0218+357, respectively
(Murphy et al. 2001).

In our calculations the $1\sigma$
confidence interval to $\Delta \mu/\mu_0$
was set as
$$
-1.5\times10^{-5} < \Delta \mu/\mu_0 < 5.7\times10^{-5}\; .
$$
For a cosmological model with $\Omega_{\rm M} = 0.3$,
$\Omega_\Lambda = 0.7$, 
and $H_0 = 72$ km~s$^{-1}$~Mpc$^{-1}$, the look-back time
for \zabs = 3.025 is 11.2 Gyr [see, e.g., equation (16)
in Carroll, Press, \& Turner 1992].
This leads to the restriction
$$
| \dot{\mu}/\mu_0 | < 5\times10^{-15}\,\, {\rm yr}^{-1}
$$
on the variation rate of $\mu_0$.

\section{Conclusions}

We have obtained a new constraint on the variation rate of
the proton-to-electron mass ratio of
$\Delta \mu/\mu_0 = (2.1\pm3.6)\times10^{-5}$ at \zabs = 3.025 toward
Q0347--3819 ($\Delta t \simeq 11$ Gyr). 
The accuracy is a factor of
3 higher as compared with the measurements in another H$_2$-bearing cloud at 
\zabs = 2.811 toward Q0528--250
(Potekhin et al. 1998).
Both measurements show no statistically significant changes of $\mu_0$
on space and time coordinates.

Since the functional dependence of the masses of proton and
electron on the fine-structure constant is unknown, we are not
able to compare directly our result with changes in $\alpha$
at the level of $\sim 0.7\times10^{-5}$ found by Webb et al. (2001).

\section*{Acknowledgments}

S.A.L. gratefully acknowledges the hospitality of the National Astronomical
Observatory of Japan, where these results have been obtained.
The authors thank A. V. Ivanchik and D. A. Varshalovich for useful discussion,
and our anonymous referee for many helpful suggestions.
The work of S.A.L. is supported in part
by the RFBR grant No.~00-02-16007.

\end{document}